\documentclass[%
 reprint,
superscriptaddress,
groupedaddress,
unsortedaddress,
 amsmath,amssymb,
 aps,
preprintnumbers
]{revtex4-2}

\usepackage{graphicx}
\usepackage{dcolumn}
\usepackage{bm}

\usepackage{hyperref}

\hypersetup{
    colorlinks=true, 
    linkcolor=white, 
    citecolor=blue,
    urlcolor=blue}


\usepackage{lipsum}
\usepackage[top= 1.3cm, bottom= 1.1cm, left= 1.8cm, right= 1.8cm]{geometry}
\usepackage{xcolor}

\begin{document}
\preprint{NuFact 2024-26}

\def\npis{$\pi^{0} \rightarrow 2\gamma~$}
\def\pis{$\pi^{0}~$}
\def\npi{$\pi^{0} \rightarrow 2\gamma$}


\title{Energy reconstruction and calibration techniques of the DUNE LArTPC}

\vspace{-4pt}
\author{\textbf{Praveen Kumar}{$^*$}, for the DUNE collaboration \vspace{1pt} }
\hypersetup{urlcolor =black}

\affiliation{Department of Physics and Astronomy, University of Sheffield, Sheffield, S3 7RH, UK}%


\thanks{\vspace{4pt}Email: praveenkumar.oblivion@gmail.com \\   Proceeding of the 25th International Workshop on Neutrinos from Accelerators (NuFact 2024), Argonne National Laboratory, IL, USA, September 2024.\vspace{-3pt}}


\begin{abstract}
The Liquid Argon Time Projection Chamber (LArTPC) technology is currently a preferred choice for neutrino experiments and beyond the standard model physics searches such as nucleon decay and dark matter. The Deep Underground Neutrino Experiment (DUNE) will employ the LArTPC technology at a large scale for physics programs, benefiting from its large target mass and excellent imaging, tracking, and particle identification capabilities. In DUNE, accurate energy reconstruction is important for precisely measuring CP violation, determining neutrino mass ordering and fully utilising the detector’s potential. The energy calibration techniques developed for the DUNE far detector (FD) horizontal drift are presented here, utilising stopping cosmic-ray muons and validating the methods with the stopping pions and protons. The study demonstrates the versatility of the calibration techniques, applicable to other LArTPC, and valid for different particles. The electromagnetic shower energy reconstruction from \npis events and the subsequent reconstruction of \pis mass are also presented. These are important calibrations which address the measurement of energy loss in the DUNE FD volume and are critical for achieving the exciting physics goals of the experiment.
\end{abstract}

\maketitle
\hypersetup{linkcolor =blue}
\vspace{-4pt}
\section{\label{sec:intro}Introduction}
\vspace{-2pt}
DUNE~\cite{dunevolume1,2dunevolume2,dunevolume3,dunevolume4} is a cutting-edge neutrino experiment that employs LArTPC technology at a large scale, due to its excellent spatial and energy resolution, as well as its remarkable particle identification capabilities. Understanding the detector’s effects is crucial for achieving the experiment's physics goals. These goals include long-baseline neutrino physics, which explores phenomena such as neutrino oscillations, CP violation, and mass hierarchy~\cite{dunesensitivity}. Additionally, DUNE will investigate beyond the Standard Model (BSM) physics, such as the search for nucleon decay. Cosmic-ray muons, a naturally occurring source capable of penetrating deep underground, serve as a ``standard candle" due to their well-understood energy loss profile and are useful in various studies, including detector calibration. Stopping muons are valuable for determining the energy scale and converting $dQ/dx$ to $dE/dx$. Muons interacting within the detector may produce both charged and neutral pions. These pions can be utilised to calibrate the DUNE FD in response to electromagnetic shower activity and the energy scale.
\vspace{-4pt}
\section{\label{sec:calibmuon}Cosmic muons in the DUNE FD}
\begin{figure}[h]
\centering
\includegraphics[trim={1cm 0.3cm 2.5cm 1cm},clip,width=.43\textwidth]{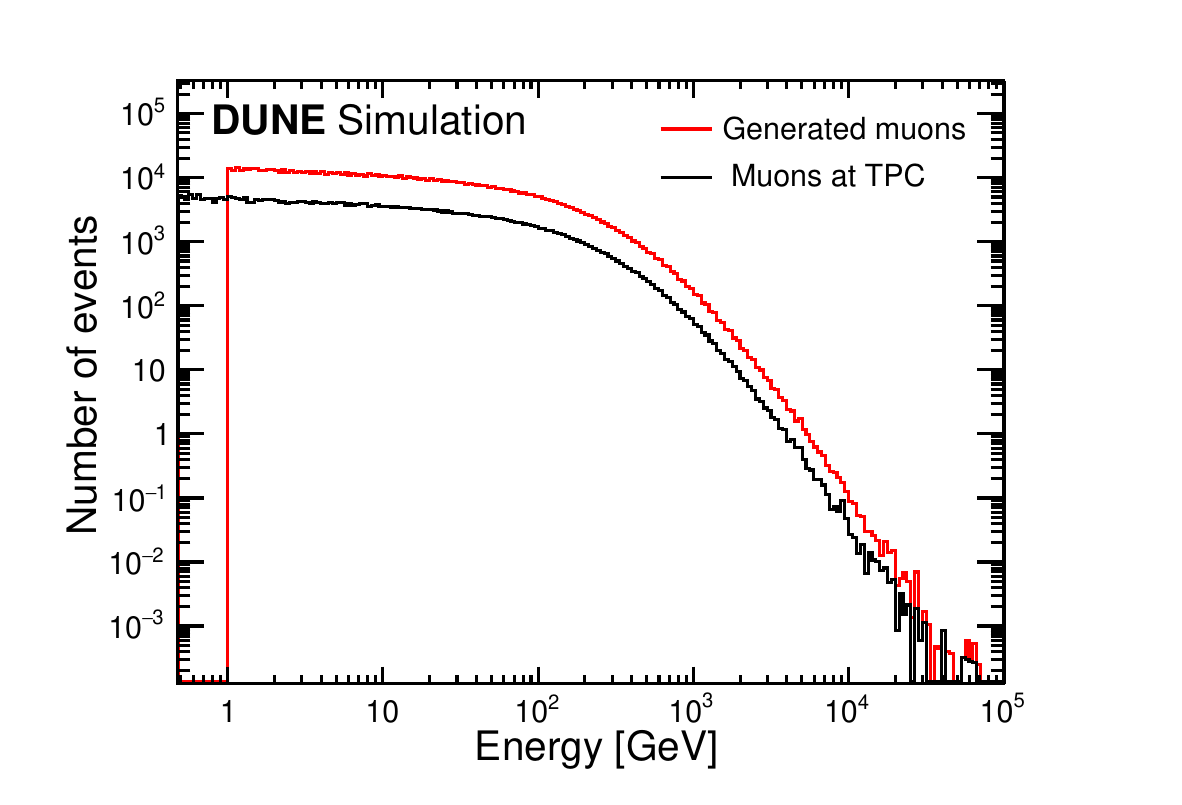}
\caption{The energy distributions of generated muons (red) and the muons as they enter the DUNE TPCs (black). A total of $1.85 \times 10^6$ simulated muon events are used.}
\label{fig1}
\vspace{-25pt}
\end{figure}

\vspace{-2pt}
MUSUN (MUon Simulation UNderground)~\cite{musun} is the muon generator useful for sampling muons in underground laboratories according to their energy and angular distributions. Using MUSUN, cosmic muons have been simulated, analysed and are used in calibrating the DUNE FD. The energy of simulated muons using MUSUN ranges from $1~\text{GeV} - 10^6~\text{GeV}$. The rate of generated cosmic muons through the box upon which muons are sampled is 0.164~Hz or 14,118 per day~\cite{lbnenote}. The muon rate inside the TPC is 0.054 Hz. A total of $1.85 \times 10^6$ muons are generated which corresponds to 131 days of data at the DUNE FD~\cite{thesispraveen}. The energy distribution of muons at the start of the active volume of the DUNE FD is shown in Fig.~\ref{fig1} (in black) and compared with generated muons (in red). 
The mean energy of muon in the DUNE FD is 284~GeV. A total of $4.76 \times 10^3$ muons enter the DUNE FD per day, with 86 muons stopping per day~\cite{thesispraveen}. 
\vspace{5pt}
\section{\label{sec:pion}{Reconstruction and identification of neutral pions}} 
\begin{figure}[h]
\centering
\includegraphics[trim={1cm .3cm 2.5cm 1cm},clip,width=.43\textwidth]{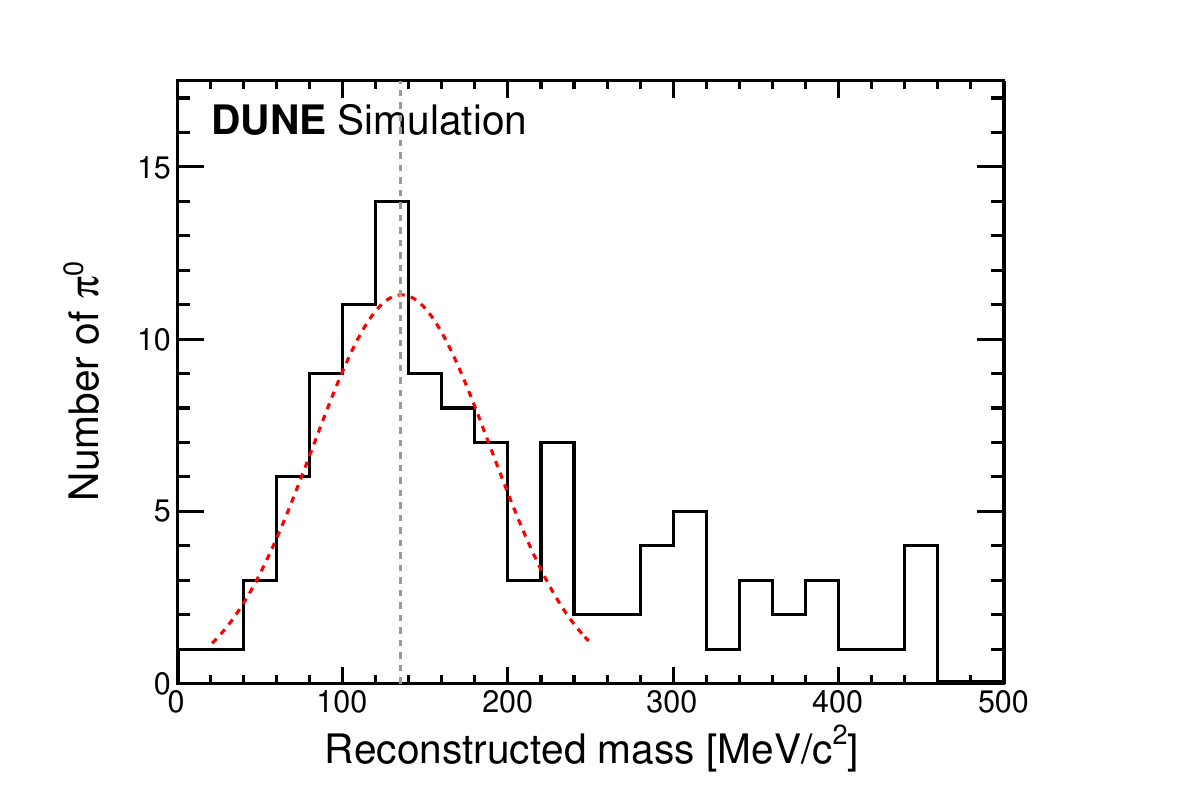}
\caption{The reconstructed mass of \pis determined from reconstructed shower pairs from \npi. The shower pairs are subjected to two requirements: an opening angle greater than $20^{\circ}$ and a minimum of 100 collection plane hits per shower. 
The red dashed line shows a Gaussian fit applied within the range [20, 250] MeV/c$^2$. The true \pis mass is shown with a vertical dashed line.}
\label{fig2}
\vspace{-15pt}
\end{figure}

Reconstruction of neutral pion events in DUNE is important for two reasons. First, the neutral pion represents a major background to the electron neutrino appearance signal. Second, the decay signature of \npis offers a valuable method for reconstructing the energy of electromagnetic showers. The true mass of the \pis can serve as a standard candle for the calibration. Studies were performed on neutral pion reconstruction produced in cosmic-ray muon events in the DUNE FD. For this study, $1.48 \times 10^6$ muon events, corresponding to 105 days of data at DUNE FD are considered. The reconstructed mass of the \pis can be determined using the reconstructed energies $(E_1,E_2)$ and opening angle ($\theta$) of the two photons from \pis decay, using the relation: $m_{\pi^{0}} = \sqrt{2 E_{1} E_{2} ( 1- \cos{\theta})}$. 
The reconstructed shower pairs are matched with photons from \npi. To select the best fraction of the sample, based on the reconstruction performance~\cite{thesispraveen}, the selections are applied to the number of reconstructed shower hits and the reconstructed opening angle of photons showers. Based on the simulation, performance cuts are applied as the reconstructed opening angle should be greater than $20^{\circ}$ and the number of reconstructed shower hits should be greater than 100. After applying these selection criteria, the number of \pis events is reduced to 156, corresponding to 105 days of data taking in a single module. The distribution of reconstructed mass is shown in Fig.~\ref{fig2}. To quantify the results, a Gaussian fit is performed within the range [20, 250] MeV/c$^2$. The reconstructed mass has a mean ($136 \pm 7$) MeV/c$^2$, which is consistent with the expected \pis mass~\cite{dunesensitivity}.

\section{\label{sec:ecalib}{Energy calibration of the DUNE FD using stopping particles}}

Stopping muons are an important calibration source for LArTPCs due to their well-understood energy loss in liquid argon. 
Determining their energy at any point along the trajectory, measured by the distance from the end of the track, known as the ‘residual range’, allows for a more detailed understanding of how the detector responds to the energy deposition by particles. The Landau-Vavilov theory accurately predicts the theoretical most probable $dE/dx$ for stopping muons in liquid argon~\cite{landau,vavilov}. The theoretical values of $dE/dx$ in liquid argon are obtained from~\cite{tablekerr}. Two techniques have been developed for the energy calibration. The first method, the modified box model~\cite{electronrecombination}, determines a calibration constant to convert ionisation charge per unit length $dQ/dx$ to particle energy loss per unit length $dE/dx$. The second method, the absolute energy scale technique~\cite{thesispraveen}, translates $dQ/dx$ to $dE/dx$ based on the comparison between theoretical $dE/dx$ and measured $dQ/dx$ as a function of residual range. 
For this study, a sample of $3.62 \times 10^5$ simulated cosmic-ray muon events, corresponding to 25 days of data at DUNE FD is considered. Muons that originate outside the TPC volume and stop inside the DUNE FD are identified as stopping muons. A total of 2,169 muons stop in the DUNE FD. To select reconstructed muons, the reconstructed track is matched with truth-stopping muons, originating outside the TPC and stopping inside the TPC volume. However, the reconstruction capability of a LArTPC is limited for tracks passing parallel to a wire or in the plane containing the drift direction and a wire. For such tracks, all the charge from the incident particle gets deposited on a single wire, leading to the poor reconstruction of the deposited charge. Tracks with certain angles with low $dQ/dx$ are removed. Additionally, due to reconstruction inefficiencies, certain particle trajectories are mistakenly reconstructed as two or more tracks; such tracks are also removed. A total of 2,140 muon tracks successfully passed all the selection criteria.
\vspace{-22pt}
\subsection{\label{sec:mbm}Modified box model} 
\vspace{-4pt}
The experimental $dE/dx$ values are derived from $dQ/dx$ values using the modified box model~\cite{electronrecombination} with the calibration constant treated as a free parameter. The modified box model equation is:
\vspace{-1pt}
\begin{eqnarray}
\left(\frac{dE}{dx}\right)_{\text{cal}} = \left(\text{exp}\left( \frac{\left(\frac{dQ}{dx}\right)_{\text{cal}} }{C_{\text{cal}}}  \frac{\beta^{'} W_{\text{ion}}}{\rho \mathcal{E} } \right) - \alpha \right) \left(\frac{\rho \mathcal{E}}{\beta^{'}}\right),~~ \label{eq:mbm}
\end{eqnarray}
where $C_{\text{cal}}$ is the calibration constant used to convert ADC values to the number of electrons, $W_{\text{ion}}$ is the work function of argon, $\mathcal{E}$ is the DUNE FD drift electric field, and  $\rho$ is LAr density at temperature 87~K. The last two parameters, $\alpha$ and $\beta^{'}$, were measured by the ArgoNeuT experiment at an electric field strength of 0.481 kV/cm~\cite{electronrecombination}. 
The calibration constant $C_{\text{cal}}$ is normalized so that the unit (“ADC$\times$tick”) corresponds to 200 electrons, where one tick corresponds to 500 ns of ADC sampling time.

\begin{figure*}[]
\begin{center}
\includegraphics[trim={0.8cm .5cm 1.5cm 1cm},clip,width=2.3in,height=1.7in]{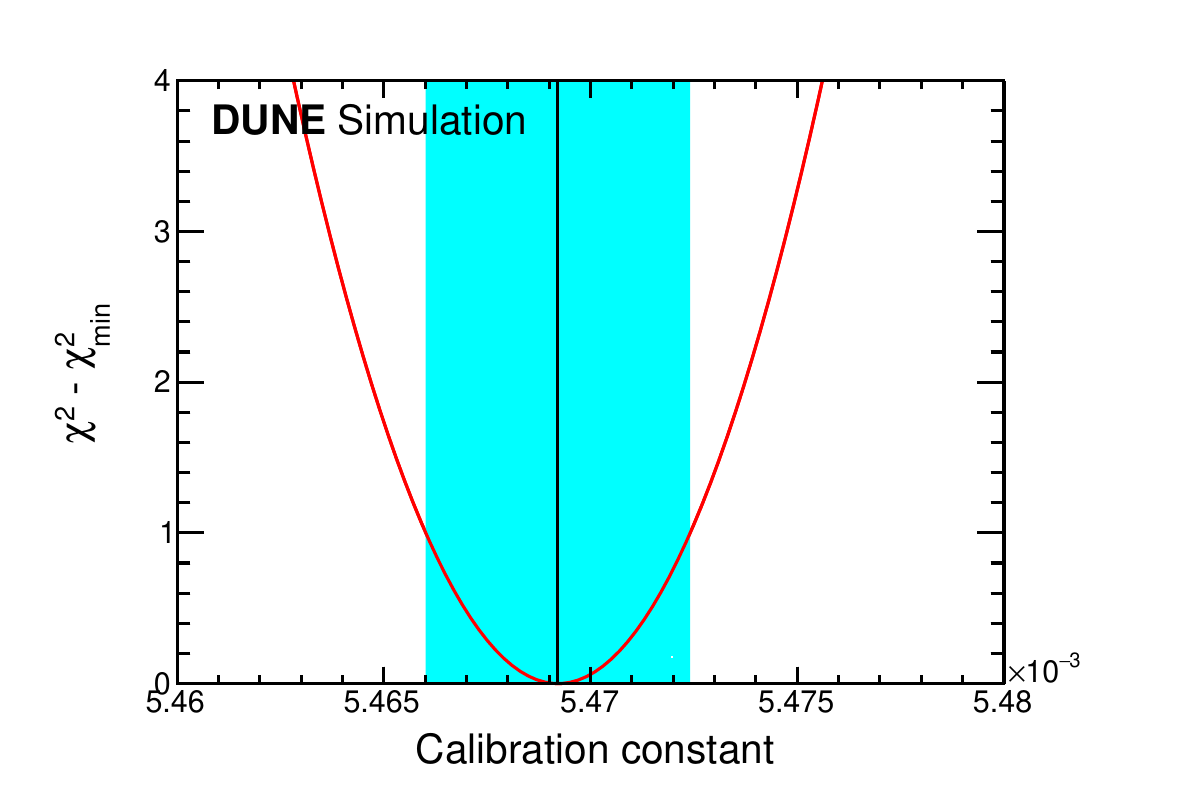}
\includegraphics[trim={0.8cm .5cm 2cm 1cm},clip,width=2.3in,height=1.7in]{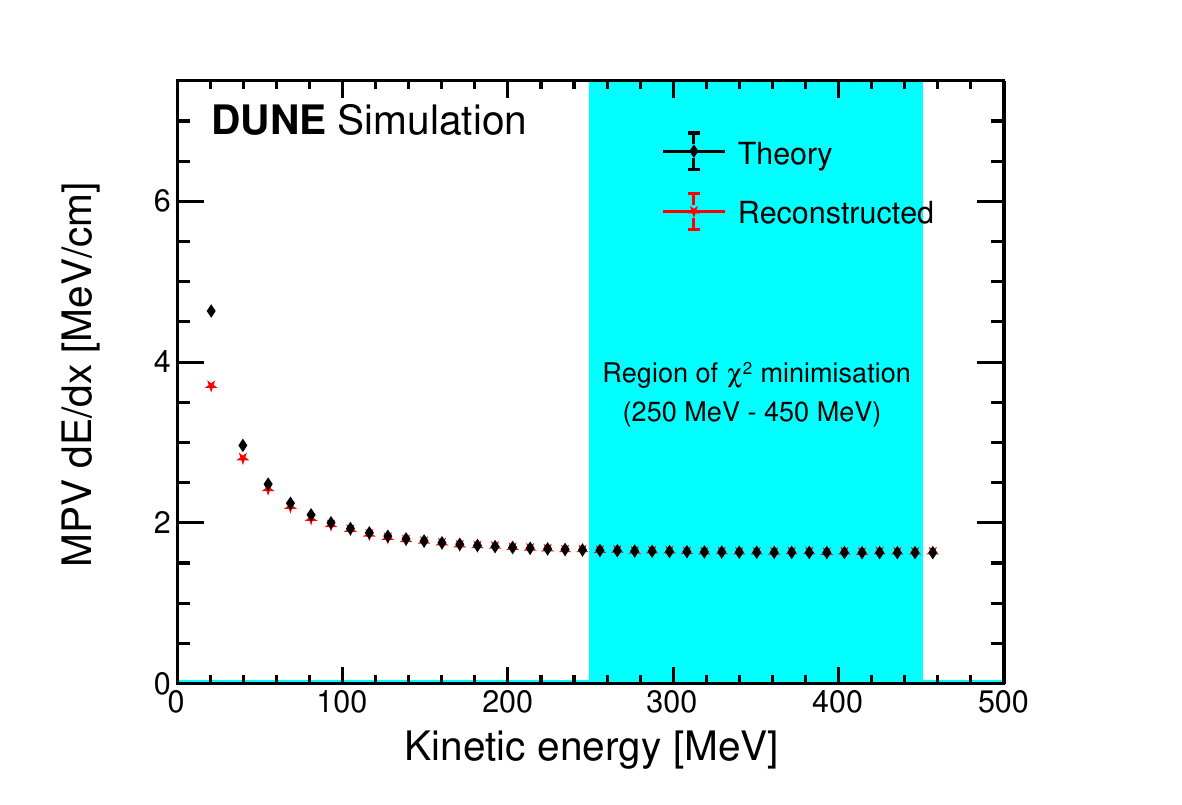}
\includegraphics[trim={0.8cm .5cm 2cm 1cm},clip,width=2.3in,height=1.7in]{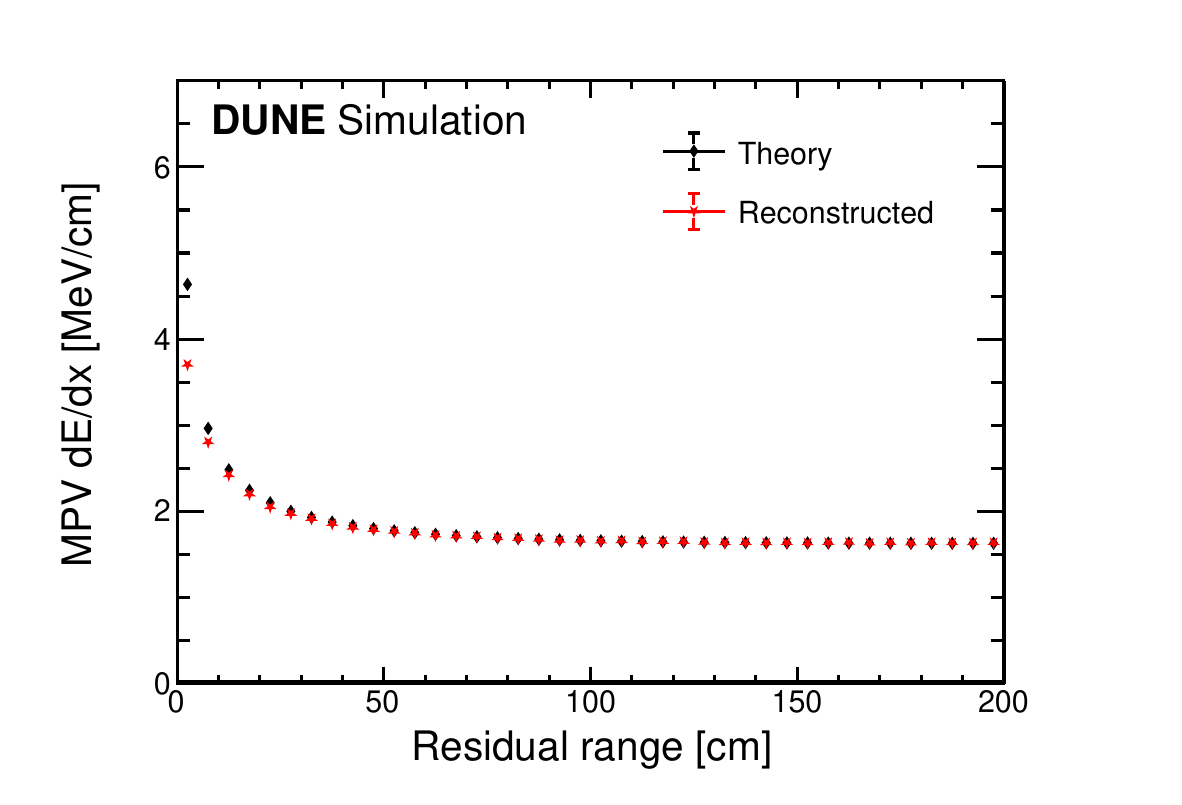}
 \vspace{-12pt}
 (a) \hspace{150 pt} (b) \hspace{150 pt} (c)
 \vspace{3pt}
\caption{Modified box model method using stopping muons: (a) $\chi^{2} - \chi_{\text{min}}^{2}$ vs. calibration constant. The black solid line represents the calibration constant at minimum $\chi^{2}$. The colour bands show the uncertainty associated with a calibration constant $C_{\text{cal}}$. (b) Comparison between the reconstructed MPV of $dE/dx$ vs. kinetic energy. (c) Comparison between the reconstructed and theoretical MPV of $dE/dx$ vs. residual range.} 
\label{fig3}
\end{center}
 \vspace{-28pt}
  \vspace{10pt}
\end{figure*}

For each selected stopping muon track, the last 200~cm from the track endpoint is divided into 5~cm bins based on residual range. In each bin, the $dE/dx$~
distribution is fitted to a Landau-Gaussian function to extract the most probable value (MPV) of $dE/dx$. 
A cubic spline interpolation based on a reference dataset is used to assign a kinetic energy at the midpoint of each bin. For bins corresponding to residual ranges of 120~--~200~cm (kinetic energies of 250~--~450~MeV), the obtained MPVs are compared to the Landau-Vavilov theoretical prediction. A $\chi{^2}$ value is calculated across a broad range of calibration constants, and the optimal calibration constant is determined as the value that minimizes $\chi{^2}$~\cite{thesispraveen}. 
Fig.~\ref{fig3}(a) shows the $\chi^{2} - \chi_{\text{min}}^{2}$ values for different calibration constants. To estimate the statistical uncertainty, $C_{\text{cal}}$ values are determined for  $\chi^{2} - \chi_{\text{min}}^{2} = 1$. The derived calibration constant value is  $(5.469 \pm 0.003) \times 10^{-3}~\text{ADC} \times \text{tick/e}$.
Fig.~\ref{fig3}(b) and \ref{fig3}(c) show the comparison between the theoretical and reconstructed MPV of $dE/dx$ as a function of kinetic energy and residual range, respectively, using the derived calibration constant.~
Most of the reconstructed values agree very well with the theoretical predictions; however, the deviation between reconstructed and theoretical values becomes noticeable at lower kinetic energy and residual range.

\begin{figure*}[ht]
\begin{center}
\includegraphics[trim={0.8cm .5cm 2cm .7cm},clip,width=2.3in,height=1.7in]{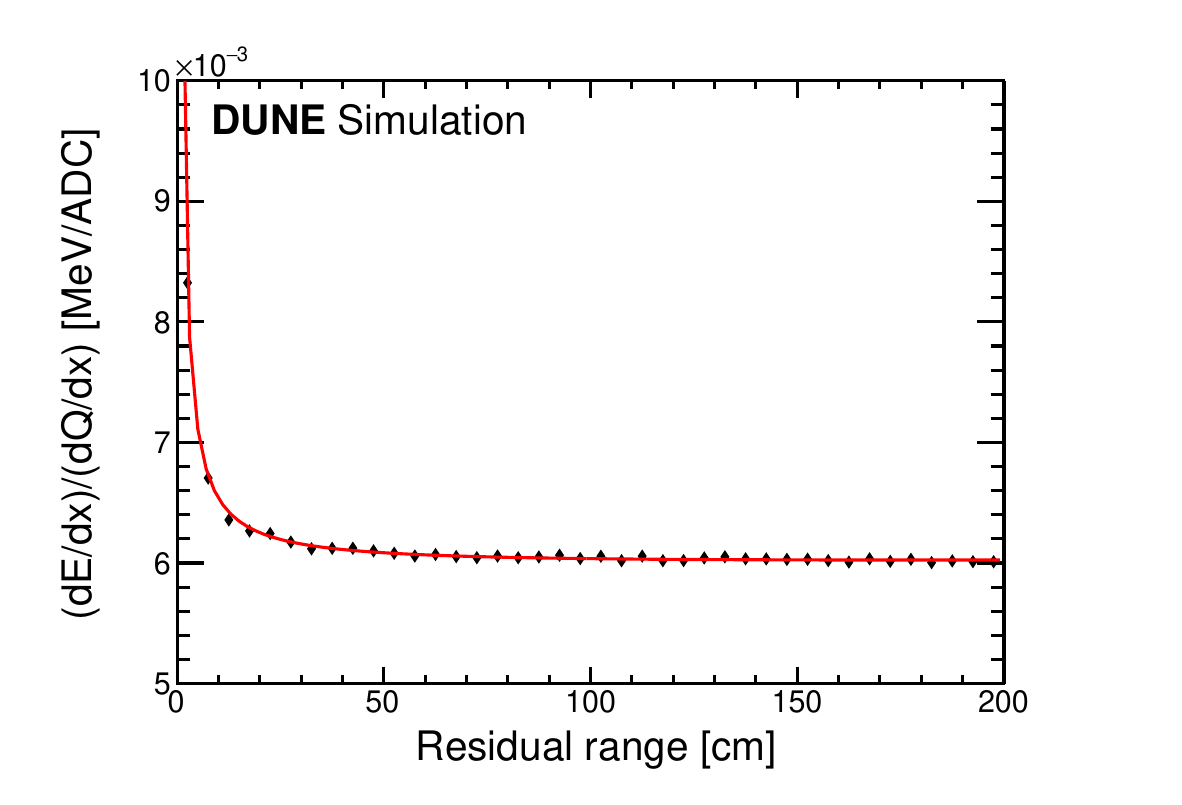}
\includegraphics[trim={0.8cm .5cm 2cm 1cm},clip,width=2.3in,height=1.7in]{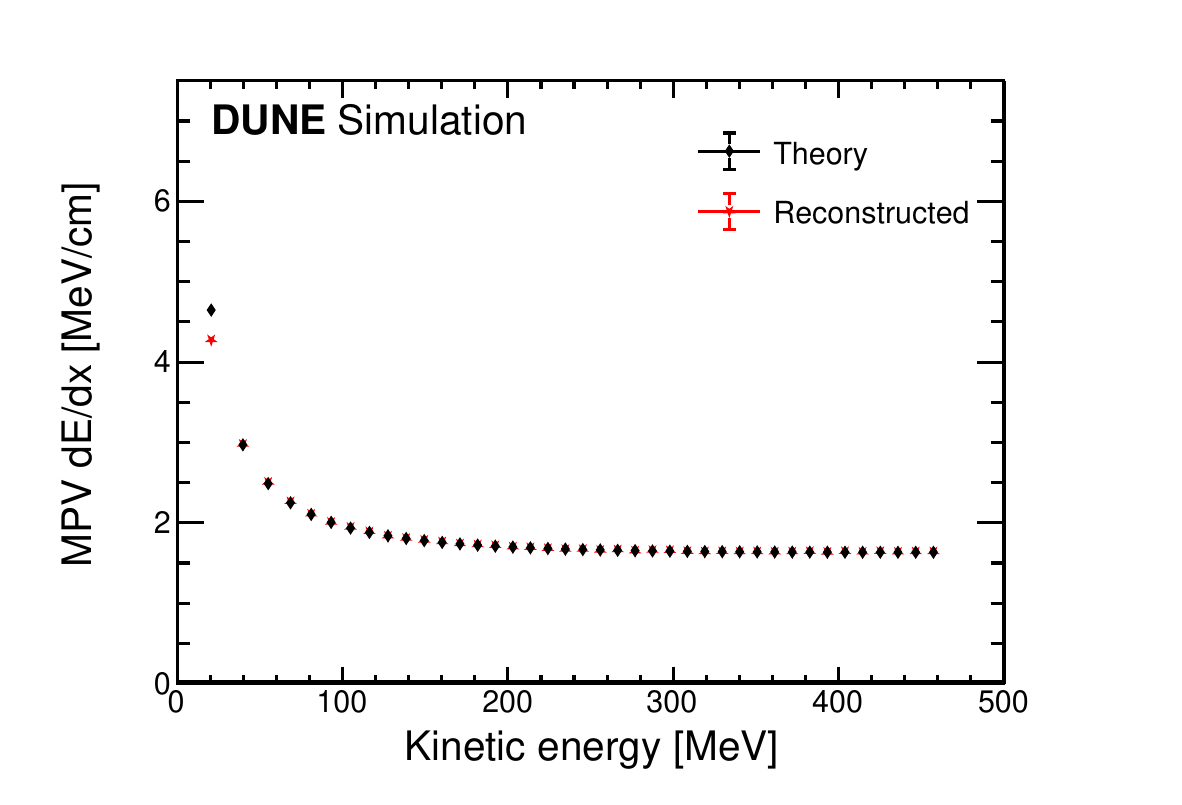}
\includegraphics[trim={0.8cm .5cm 2cm 1cm},clip,width=2.3in,height=1.7in]{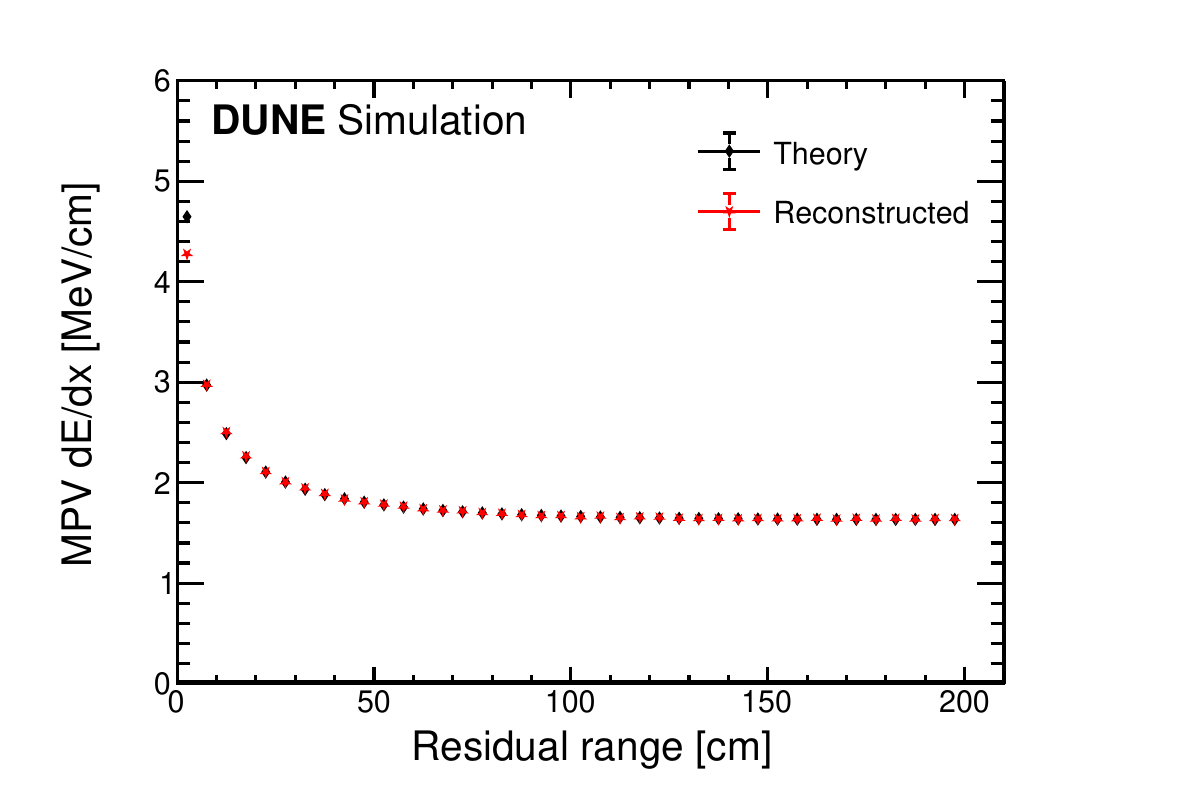}
 \vspace{-12pt}
 (a) \hspace{150 pt} (b) \hspace{150 pt} (c)
 \vspace{3pt}
\caption{Absolute energy scale method using stopping muons: (a) The ratio of theoretical $dE/dx$ to reconstructed $dQ/dx$ of stopping muons as a function of residual range, fitted with the function described in Equation~\ref{eq:fitfn}. (b) The reconstructed MPV of $dE/dx$ vs. kinetic energy compared with the theoretical MPV of $dE/dx$. 
(c) The reconstructed MPV of $dE/dx$ vs. residual range, compared with the theoretical MPV of $dE/dx$. The reconstructed $dE/dx$ agrees well with the theoretical prediction, even at low kinetic energy and residual range. The reconstructed (red) and theoretical (black) points overlap for most values.}
\label{fig4}
\end{center}
 \vspace{-15pt}
\end{figure*}

\begin{figure*}
\begin{center}
\includegraphics[trim={0.8cm 0.5cm 2cm 1cm},clip,width=2.32in,height=1.7in]{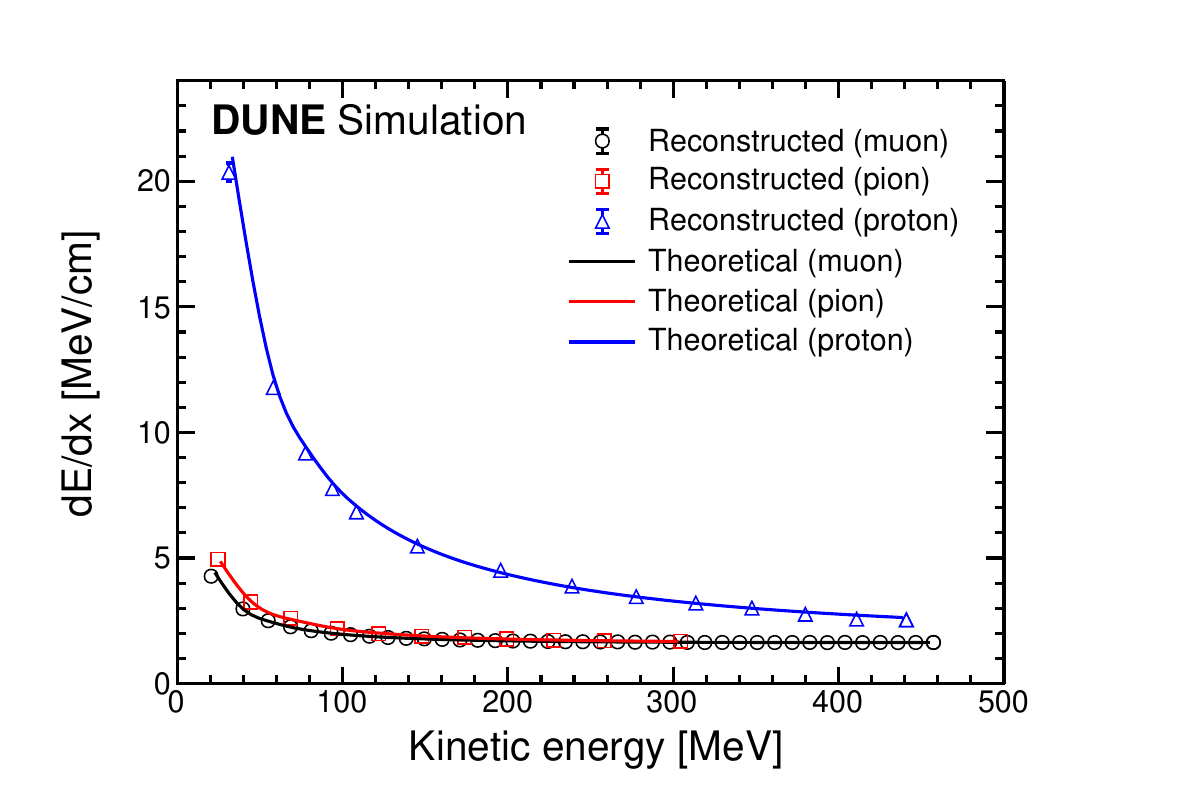}
\includegraphics[trim={0.8cm 0.5cm 2cm 1cm},clip,width=2.32in,height=1.7in]{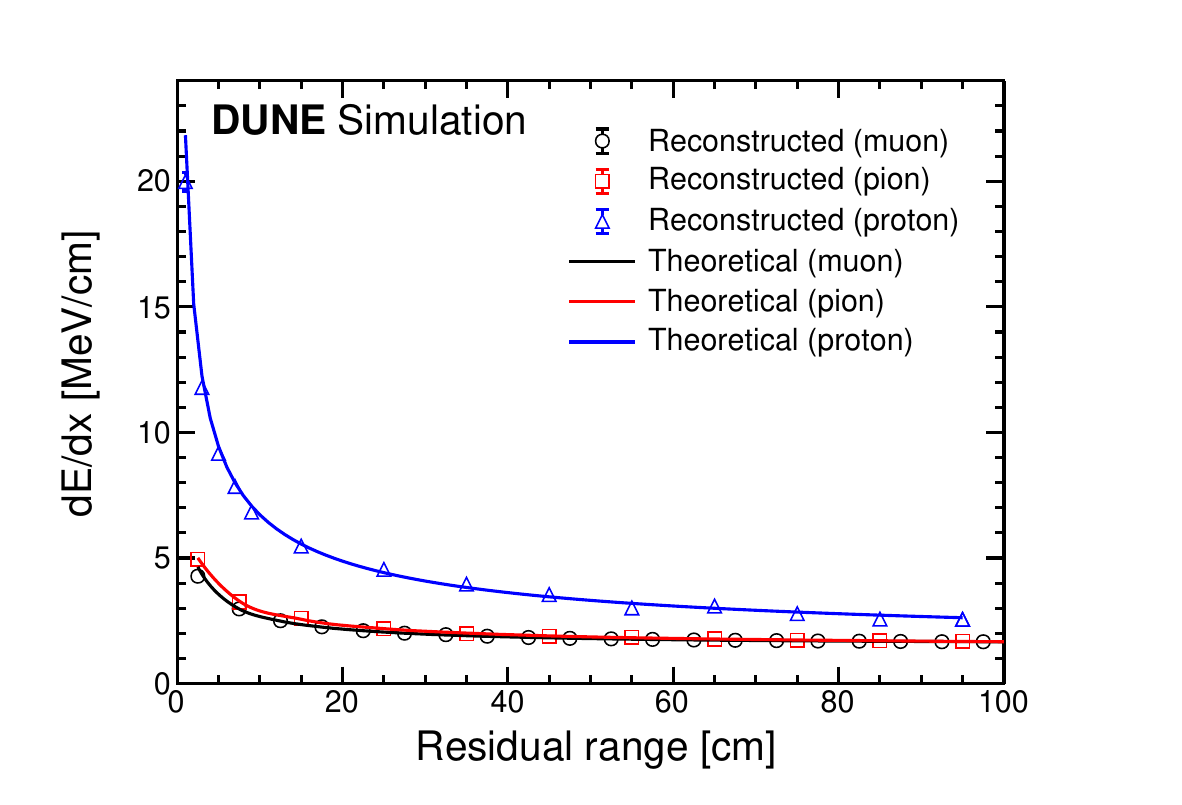} 
\includegraphics[trim={0.8cm 0.5cm 2cm 1cm},clip,width=2.32in,height=1.7in]{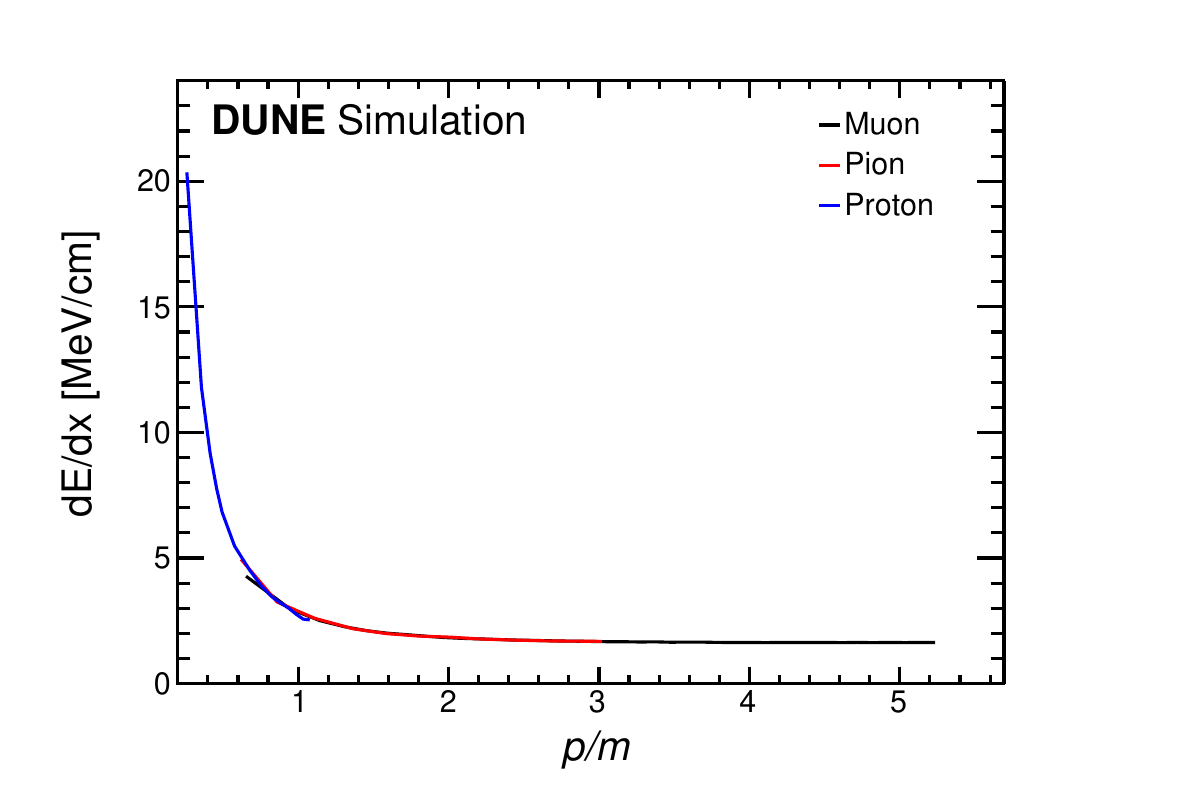}
 \vspace{-12pt}
 (a)\hspace{150 pt} (b) \hspace{150 pt} (c)
 \vspace{3pt}
\caption{Comparision of $dE/dx$ reconstruction of stopping particles: (a) The reconstructed MPV of $dE/dx$ as a function of kinetic energy for muons, pions, and protons compared with theoretical values. (b) The reconstructed MPV of $dE/dx$ as a function of residual range for muons, pions, and protons compared with theoretical values. Good agreement is observed for reconstructed and theoretical prediction based on the Landau-Vavilov theory~\cite{landau,vavilov}. (c) The reconstructed MPV of $dE/dx$ as a function of $p/m$. The $dE/dx$ value is almost the same for muons, pions, and protons at the same $p/m$.
}
\label{fig5}
\end{center}
 \vspace{-15pt}
\end{figure*}

\vspace{-24pt}
\subsection{\label{sec:aes}Absolute energy scale}
\vspace{-4pt}
A technique is developed based on the residual range, independent of any specific model for translating $dQ/dx$ to $dE/dx$~\cite{thesispraveen}. 
For each stopping muon track, the last 200~cm (or less) from the track end is divided into 5~cm bins based on residual range. The reconstructed $dQ/dx$ distribution in each bin is fitted with a Landau function convolved with a Gaussian to determine its MPV. Using Landau-Vavilov theory, the theoretical MPV of $dE/dx$ is calculated for each bin. The ratio of theoretical $dE/dx$ to reconstructed $dQ/dx$ is then plotted against residual range and fitted with an empirical function: 
\begin{equation}
f(r) = p_0~+~p_1\times1/r~+~p_2\times r, ~~\label{eq:fitfn}
\end{equation}
as shown in Fig.~\ref{fig4}(a) and thus a relationship between reconstructed $dE/dx$ and $dQ/dx$ is obtained:
\begin{equation}
\frac{\text{MPV}(dE/dx)_{reconstructed}} {\text{MPV}(dQ/dx)_{reconstructed}} = p_0~+~p_1\times1/r~+~p_2\times r,~~\label{eq:fitfn1}
\end{equation}
where $r$ represents residual range in cm and $p_0, p_1$ and  $p_2$ are the fitting parameters. The fitted parameter values are: $p_0 = (5.962 \pm 0.007) \times 10^{-3}$, $p_1 = (5.731 \pm 0.087) \times 10^{-3}$ and $p_3 = (1.613 \pm 0.499) \times 10^{-7}$. 
The reconstructed $dQ/dx$ is converted to reconstructed $dE/dx$ using Equation~\ref{eq:fitfn1}. Figs.~\ref{fig4}(b) and \ref{fig4}(c) show the reconstructed $dE/dx$ and compare with theoretical $dE/dx$ as functions of kinetic energy and residual range, respectively, demonstrating improvement in low residual ranges over the modified box model. A key advantage of the absolute energy scale lies in its independence from external models or parameters derived from other experiments. 
This emphasises the efficiency and autonomy of the absolute energy scale method, positioning it as a complementary method for $dE/dx$ calibration in the DUNE experiment.

After establishing the absolute energy scale for stopping muons, $dE/dx$ calibration is performed with pions ($\pi^+$ and $\pi^-$), and protons. The studies utilise the same cosmic-ray muon sample as in the stopping muon analysis. Pions produced by the cosmic-ray muons interacting in the liquid argon were selected for this study, resulting in a total of $9.30 \times 10^3$ pions and $5.34 \times 10^4$ protons. 
Using the same absolute energy scale method as for the stopping muons, the reconstructed $dE/dx$ for pions and protons was obtained and compared with the theoretical values. Fig.~\ref{fig5}(a) and \ref{fig5}(b) show the comparison of reconstructed and theoretical $dE/dx$ as a function of the kinetic energy and residual range for muons, pions and protons. The energy loss dependence on a particle mass in a medium can be inferred from the Bethe-Bloch formula~\cite{bethebloch}. These particles show good agreement of reconstructed $dE/dx$ with the theoretical prediction. The $dE/dx$ value is the same for muons, pions, and protons at the same momentum-to-mass ratio ($p/m$) within the medium, as described by the Bethe-Bloch formula~\cite{bethebloch}. Fig.~\ref{fig5}(c) illustrates the reconstructed MPV of $dE/dx$ as a function of the $p/m$ for stopping muons, pions and protons. These particles demonstrate the same $dE/dx$ at the same $p/m$ ratio (in natural units) in liquid argon. This, in turn, validates the calibration process using the absolute energy scale method employed in the conversion of $dQ/dx$ to $dE/dx$ for muons, pions and protons. The uniformity in the $p/m$ behaviour across all three particles confirms the accurate conversion of $dQ/dx$ to $dE/dx$ using the absolute energy scale method.
\vspace{.3cm}
\vspace{-11.5pt}
\section{\label{sec:sum}{Conclusions}}
Cosmic-ray muon events are simulated with the MUSUN generator for the DUNE FD and are used for energy reconstruction and calibration. The decay signature of $\pi^{0}\rightarrow2\gamma$ provides a valuable tool for calibrating the energy of electromagnetic showers. The reconstructed \pis mass is determined to be $(136 \pm 7)~\text{MeV/c}^2$, consistent with the expected \pis mass. Several areas for improvement remain, particularly in shower energy reconstruction. Two energy calibration techniques were developed using stopping particles for the DUNE FD. Using the modified box model, a calibration constant of $C_{\text{cal}} = (5.469 \pm 0.003) \times 10^{-3}~\text{ADC} \times \text{tick/e}$ is derived for stopping muons. In addition, an absolute energy scale technique was developed, enabling accurate conversion of $dQ/dx$ to $dE/dx$ for stopping muons. This absolute energy scale was applied to other particles, such as pions and protons, demonstrating good agreement between the reconstructed most probable $dE/dx$ and the theoretical prediction based on the Landau-Vavilov theory. 
The $dE/dx$ value is the same for muons, pions, and protons at the same $p/m$ highlights the independence of the momentum-to-mass ratio from particle type, validating the calibration method. This study highlights the versatility of the calibration technique, demonstrating its applicability to other LArTPC and its validity for various particle types.


\nocite{}
\hypersetup{urlcolor =blue}
\bibliography{Proceeding_Nufact}
\pdfinfo{
  /Creator (Praveen Kumar)
  /Author (Praveen Kumar)
  /Keywords (Paper)}

\end{document}